\documentclass[a4paper]{article}
\usepackage[utf8]{inputenc} 
\usepackage[T2A]{fontenc}
\usepackage{hyperref}
\usepackage{amssymb,amsmath}
\usepackage{floatrow}
\usepackage{layout}
\usepackage{amssymb} 
\usepackage{multirow}
\usepackage{caption}
\usepackage{authblk}
 \usepackage{indentfirst}
\usepackage{geometry}
\usepackage[super,compress]{cite}
\usepackage{graphicx}

\title{Induced gravity and cosmological principle}
\author{V. A. Berezin$^1$, I. D. Ivanova$^2$, A. E. Kuprina$^3$}

\date{ \small
	$^1$Institute for Nuclear Research of the Russian Academy of Sciences, \\
60th October Anniversary Prospect 7a, 117312 Moscow, Russia \\ \texttt{berezin@inr.ac.ru}\\%
	$^2$Institute for Nuclear Research of the Russian Academy of Sciences, \\
60th October Anniversary Prospect 7a, 117312 Moscow, Russia \\ \texttt{pc\textunderscore mouse@mail.ru}\\%
$^3$National Research University "Higher School of Economics", \\
20, Myasnitskaya Street 101000, Moscow, Russia \\ \texttt{stazz2016@gmail.com}\\[2ex]%
}

\begin{document}
\maketitle

\begin{abstract}
The phenomenological description of the cosmological particle production in the framework of the induced gravity is investigated. It appears that the source terms with the particle number density in the creation law can be interpreted as the invisible part of the Universe. It is shown that there is a gauge that restores the General Relativity in which our model is equivalent to the $f(R)$-gravity for $f \propto R^{\frac{3}{2}}$.
\end{abstract}

\section{Introduction} 
\paragraph{} The identity of the gravitational and inertial masses discovered by Galileo Galilei \cite{Galileo} was developed by Albert Einstein \cite{AE1} up to the equivalence principle and became the milestone in constructing General Relativity \cite{AE2}. David Hilbert showed that Einstein equations can be obtained from the least action principle using the simplest possible geometrical invariant (the curvature scalar) and Riemannian geometry \cite{Hilbert}. Moreover, he introduced new definition for the energy-momentum tensor of the matter fields (the so-called metrical energy-momentum tensor) which can be naturally considered as a source of any geometrical theory of gravitation \cite{Hilbert}\cite{LL}.
\par Is the gravitation fundamental interaction field, like the electromagnetic electroweak and strong ones?
\par The most radical view was presented by A.D. Sakharov \cite{Sakh}. According to him, the gravitation is not the fundamental field existing by itself, but it is just the tension of the quantum vacua of all other quantized fields. Now such an approach is known as the induced gravity. In this paper we adopt this point of view.
\par One of the most intriguing features of quantum field theory is the possibility of the creation of particles. In the Minkowski (flat) space-time one needs to form strong (or with strong gradients) classical fields. But in the curved space-times the gravitation leads to the polarization of the quantum vacua, and the particle creation may occur even in the absence of the classical fields. Evidently, such a situation is very much welcome both in cosmology and black hole physics. In the present paper we are interested in cosmology only.
\par The cosmological particle creation was intensively studied in the early 70's of the last century by Z. Parker and S. Fulling, Y.B. Zel'dovich and A.A. Starobinsky, A.A. Grib, V.M. Mostepanenko and S.G. Mamaev and many other groups in different countries \cite{Cr1,Alstar1,Cr3}. All of them was developing the theory of the quantum scalar field on the given curved background cosmological space-times and the process of creation of the scalar particles from the vacuum. Namely, they considered the homogeneons and slightly anisotropic cosmological models. In 1977 Zel'dovich and Starobinsky published short, but remarkable paper \cite{Alstar1} where they showed that the rate of the particle creation (i.e., the number of particles produced per unit time interval and unit volume) is proportional to the square of the Weyl tensor. The main qualitative results were the discovery of the essential importance of the so called trace anomaly of the quantum scalar field on the curved background and the absence of the particle production the homogeneous and isotropic cosmological models. It is the trace anomaly that provides us with the necessary interactions between the quantum vacuum and the geometry. Note, that in the Minkowski space-time the trace anomaly is absent. The second result is due to the fact that for any homogeneous and isotropic cosmological model the Weyl tensor is identically zero.
\par We would like to emphasize here that the problem of the cosmological particle creation was not completely solved, because the back reaction of the already created particles and, what seems much more important, the very process of their creation on the space-time metric was not taken into account. The attempts to combine the essentially quantum process of the particle creation with the essentially classical geometry reveals the huge problem. The matter is that in order to solve the quantum problem one needs to impose some boundary conditions, and to know the boundary conditions one have to solve the gravitational equations. But, the source of the gravitation is just the quantum field averaged over some vacuum state which remains unknown prior to solving these classical equations determining the global geometry. How to break this closed circle?
\par We propose to resolve this difficulty phenomenologically. First, we will describe the already created particles by the hydrodynamics. In the usual hydrodynamics the total number of particles is assumed to be conserved. Therefore, we need the unusual hydrodynamics \cite{Ber1}, with explicit creation law equal to the rate of particle production. This creation law should include the pure gravitational (=geometrical) part responsible for the polarization of the vacuum, the scalar field which we assume to produce particles, the term(s) describing the interaction of this scalar field with the geometry, and the already created particles because they are nothing more but the on-shell quanta of that scalar field.
\par Our phenomenology opens the possibility for describing the vacuum polarization, it is sufficient to put the creation law zero, but not identically zero. Also, we will show that the accounting for the back reaction changes the whole situation drastically. Namely, as it will be shown, the second result of the previous investigators is no more valid. The particle creation becomes possible even in the homogeneous and isotropic models. Moreover, it is this process that may be responsible for the appearance of the gravitating dark matter. There exists also the room for the dark energy.
\section{Mathematical tools}
\par We will be working in the framework of Riemannian geometry and using the units with $\hbar=c=1$. The interval $ds$ between the nearby points of the space-time is determined by the metric tensor $g_{\mu\nu}(x)$
$$ds^2=g_{\mu\nu}dx^\mu dx^\nu$$
with the signature $(+---)$, greek indices run 0, 1, 2, 3. The Riemann curvature tensor is defined as follows
$$R^\mu_{\nu\lambda\sigma}=\frac{\partial\Gamma^\mu_{\nu\sigma}}{\partial x^\lambda}-\frac{\partial\Gamma^\mu_{\nu\lambda}}{\partial x^\sigma}+\Gamma^\mu_{\kappa\lambda}\Gamma^\kappa_{\nu\sigma}-\Gamma^\mu_{\kappa\sigma}\Gamma^\kappa_{\nu\lambda},$$
where the connections $\Gamma^\mu_{\nu\lambda}$ are just the Christoffel symbols
$$\Gamma^\mu_{\nu\lambda}=\frac{1}{2}g^{\mu\sigma}(g_{\sigma\nu,\lambda}+g_{\sigma\lambda,\nu}-g_{\nu\lambda,\sigma}),$$
here $g^{\mu\sigma}$ is the inverse metric tensor $(g^{\mu\sigma}g_{\sigma\nu}=\delta^\mu_\nu)$, and comma "," denotes the partial derivative. The convolutions of the curvative tensor give the Ricci tensor $R_{\mu\nu}$ and the curvature scalar R:
$$R_{\mu\nu}=R^\sigma_{\mu\sigma\nu}, R=g^{\mu\nu}R_{\mu\nu}=R^\sigma_\sigma,$$
the indices are raising and lowering by the metric tensor and its inverse. We will also need the so called Weyl tensor which is defined as the completely traceless part of the Riemann curvature tensor,
$$C_{\mu\nu\lambda\sigma}=R_{\mu\nu\lambda\sigma}-\frac{1}{2}R_{\mu\lambda}g_{\nu\sigma}+\frac{1}{2}R_{\mu\sigma}g_{\nu\lambda}-\frac{1}{2}R_{\nu\sigma}g_{\mu\lambda}+\frac{1}{2}R_{\nu\lambda}g_{\mu\sigma}+\frac{1}{6}R(g_{\mu\lambda}g_{\nu\sigma}-g_{\mu\sigma}g_{\nu\lambda}).$$
\par In what follows, the very important role will be played by the local conformal transformation defined as follows:
$$ds^2=\Omega^2d\hat{s}^2 = \Omega^2(x)\hat{g}_{\mu\nu}dx^\mu dx^\nu,$$
where ``\^{}'' means "transformed", and $\Omega(x)$ is called "the conformal factor". We would like to emphasize that such a transformation does not touch the coordinate system (the frame of reference).
\par Note, that the Weyl tensor $C^\mu_{\nu\lambda\sigma}$ is conformal invariant,
$$C^\mu_{\nu\lambda\sigma}=\hat{C}^\mu_{\nu\lambda\sigma}.$$
\par In the present paper we are interested in the cosmological solutions. By cosmology we understand the homogeneous and isotropic models with the Robertson-Walker metric
$$
ds^2=dt^2-a^2(t)dl^2,
$$
$$l^2=\frac{dr^2}{1-kr^2}+r^2(d\theta^2+\sin^2\theta d\varphi^2), \quad (k=0,\pm1),
$$
where $a(t)$ is the scalar factor, $k=1$ stands for the closed model, while $k=0$ for the spatially flat, and $k=-1$ for the open ones. The temporal coordinate t is called the cosmological time. The so called conformal time $\eta$ is introduced by the following transformation:
$$ds^2=a^2(t)\left(\frac{dt^2}{a^2}-dl^2\right)=\widetilde{a}^2(\eta)(d\eta^2-dl^2),$$
further we will omit the superscript "tilde". Note, that the differential of the cosmological time $dt$ is not conformal invariant. Indeed,
$$ds^2=dt^2-a^2(t)dl^2=\Omega^2(t)\left(\frac{dt^2}{\Omega^2(t)}-\hat{a}^2(t)dl^2\right)=\Omega^2(t)(d\hat{t}^2-\hat{a}^2(t)d\hat{l}^2),$$
so,
$$d\hat{l}=dl, \quad d\hat{t}=\frac{dt}{|\Omega(t)|}.$$
On the contrary, the differential of the conformal time, $d\eta$, is obviously, conformal invariant, and the scale factor $a(\eta)$ serve as the conformal factor when one starts from $a=1$.
\par For any homogeneous and isotropic cosmological model the Weyl tensor is identically zero,
$$C^\mu_{\nu\lambda\sigma}\equiv 0.$$

\section{Construction of the Lagrangian}
\par It was claimed in Introduction that we adopt the conception of the induced gravity. Therefore, the total action integral, $S_{tot}$, becomes just the matter action integral, $S_m$:
$$S_{tot} = S_m.$$
\par It was also claimed that created particles will be described by hydrodynamics. The simplest possible choice is the perfect fluid. It is well known that the usual hydrodynamics can be described in two different ways - by Lagrangian and by Eulerian pictures. The first of them requires the "eternal" existence of particle's wordlines. Since we are interested in the particle creation (and annihilation), it is not good for us. Thus, we have to use the Eulerian picture.
\par Let us start with the usual hydrodynamics postulated the particle number conservation law. The corresponding action integral in the Eulerian variables was elaborated by J.R. Ray \cite{Ray}:
\begin{eqnarray}
S_{\rm m}&=& -\!\int\!\varepsilon(X,n)\sqrt{-g}\,d^4x + \int\!\lambda_0(u_\mu u^\mu-1)\sqrt{-g}\,d^4x+\nonumber\\
&&+\int\!\lambda_1(n u^\mu)_{;\mu}\sqrt{-g}\,d^4x+ \int\!\lambda_2 X_{,\mu}u^\mu\sqrt{-g}\,d^4x.
\end{eqnarray}
Here the semicolon ";" denotes the covariant derivative. The dynamical variables are the particle number density $n(x)$, the vector field $u^\mu(x)$ and the auxiliary variable $X(x)$, while $\lambda_0(x)$, $\lambda_1(x)$ and $\lambda_2(x)$ are Lagrange multipliers. The energy density $\varepsilon$ depends only on $n$ and $X$, and the Lagrange multipliers provide us with three constraints:
$$u^\sigma u_\sigma - 1 = 0$$
makes the vector field $u^\mu$ similar to the four-velocity of particles,
$$(nu^\sigma)_{;\sigma}=0$$
manifests the particle conservation law, and
$$X_{,\mu}u^\mu=0$$
links the vector field $u^\mu$ to the particle trajectories and enumerates them.
\par In the homogeneous and isotropic cosmological models the Robertson-Walker frame of reference (the coordinate system) is supposed to be comoving. So, $u^t=u_t=1, u^i=0$ $(i=1,2,3)$ if one uses the cosmological time $t$, and $u^\eta=\frac{1}{a}$, $u_\eta=a$, $u^i=0$, if one uses the conformal time $\eta$. In the latter case the surface $a=0$ is null, not space-like as in the case of the cosmological time.
\par The advantage of the above form of the Lagrangian for the perfect fluid is that the particle number conservation condition enters explicitly. Therefore, it is straightforward to replace it by the creation law,
$$
(nu^\mu)_{;\mu}=\Phi[\rm inv].
$$
The creation function $\Phi[\rm inv]$ consists of some invariants constructed form metric tensor and some fields whose quanta are supposed to be created. Fortunately, it is possible to be more precise. Let us consider the left hand side of the above equation under the local conformal transformation. Evidently, by definitions,
$$n=\frac{\hat n}{\Omega^3},  \quad u^\mu=\frac{\hat u^\mu}{\Omega},  \quad \sqrt{-g}=\Omega^4 \sqrt{-\hat g}.$$
Using the famous relation $l^\mu_\mu=\frac{(l^\mu\sqrt{-g})}{\sqrt{-g}},\mu$ valid in Riemannian geometry for any vector $l^\mu$, one readily obtains:
$$
(nu^\mu)_{;\mu} = \frac{(nu^\mu\sqrt{-g})_{,\mu}}{\sqrt{-g}}=\frac{\left(\frac{\hat{n}}{\Omega^3}\frac{\hat{u}^\mu}{\Omega}\Omega^4\sqrt{-\hat{g}}\right)_{,\mu}}{\sqrt{-g}} = \frac{(\hat{n}\hat{u}^\mu\sqrt{-\hat{g}})_{,\mu}}{\sqrt{-g}}.
$$
Therefore $(nu^\mu)_{;\mu}\sqrt{-g}$ is conformal invariant \cite{Ber4}. Hence, $\Phi\sqrt{-g}$ must be conformal invariant.
\par The creation law $\Phi$ should contain a pure geometrical contribution which causes the vacuum polarization. There exists only one candidate (up to quadratic in curvatures) - the square of the Weyl tensor, $C^2=C_{\mu\nu\lambda\sigma}C^{\mu\nu\lambda\sigma}$:
$$\Phi=\alpha C^2+...$$
Then, here we will consider only the particle production by some scalar field $\varphi$. Assuming its usual transformation under the local conformal transformation, namely,
$$\varphi=\frac{\hat{\varphi}}{\Omega},$$
one has at hand the following very famous suitable combination
$$\varphi\square\varphi - \frac{1}{6}\varphi^2R+\Lambda\varphi^4,$$
here $\square$ stands for d'Alambertian, $\square\varphi=\varphi^\sigma_{;\sigma}$ $(\varphi^\sigma = g^{\sigma\kappa}\varphi_\kappa = g^{\sigma\kappa}\varphi_{,\kappa})$, and $R$ is the curvature scalar. Thus, the creation law becomes
$$
\Phi=\alpha \, C^2+\beta\, \left( \varphi \square \varphi -\frac{1}{6}\, \varphi ^{2}\, R+\Lambda \, \varphi ^{4} \right)+...
$$
We consider the scalar field $\varphi$ as describing the vacuum fluctuation and the germ for particle creation, while the classical scalar field is concentrated in the already created particles. This means that there should be the contribution to the creation law depending on the particle number denote by $\varepsilon_1(\varphi,n)$. Thus, eventually,
$$
\Phi=\alpha \, C^2+\beta\, \left( \varphi \square \varphi -\frac{1}{6}\, \varphi ^{2}\, R+\Lambda \, \varphi ^{4} \right)+\varepsilon_1(\varphi,n).
$$
\par Some remarks are in order. We have already mentioned about one of the main results of the previous investigators, namely, the absence of the particle production in the homogeneous and isotropic cosmological models. Now, it is becoming clear that this is entirely due to neglecting of the back reaction of the scalar field fluctuation on the space-time metric and zero value of the Weyl tensor. Evidently, this is no more true in our phenomenological model. The particle can be created in homogeneous and isotropic manner. Its gravitational influence may appear much more stronger than that of ordinary (visible) matter created by the inhomogeneous and anisotropic fluctuations considered by V. F. Mukhanov and G. V. Chibisov \cite{Muhanov Chibisov}. Moreover they may be thought of as the dark matter. At the same time, the $\varphi^4$- V.F. Mukhanov, G.V. Chibisov term in the creation law is the good candidate for the quintessence. Note also the appearance of the $\varphi^2R$-term which can be interpreted as the interaction between the vacuum fluctuations and the space-time geometry.
\par At the end of this section we would like to investigate the structure of the energy density $\varepsilon$ and the term $\varepsilon_1$ in the creation law. To do this, let us consider their behavior under the local conformal transformation. Due to the adopted induced gravity point of view, $S_{tot}=S_m$. Surely, the action integral does not need to be conformal invariant, but its variation does because this transformation does not touch the coordinate system (and hence, the limits of the integration). Indeed,
$$\frac{\delta S}{\delta \Omega}=\frac{\delta S}{\delta \psi}\frac{\delta \psi}{\delta \Omega}\equiv0.$$
($\psi$ is the collective dynamical variable). It is easy to prove that the terns with the Lagrange multipliers in our action $S_m$ are conformal invariant. So, we have to check only the first term,
$$\frac{\delta}{\delta\Omega}\int \varepsilon (X,\varphi, n) \sqrt{-g}\,  d^{4}x.$$
Since the enumeration of the trajectories, $X$, is not touched by the conformal transformation, and $n=\frac{\hat n}{\Omega^3},$ $u^\mu=\frac{\hat u^\mu}{\Omega},$ $\sqrt{-g}=\Omega^4 \sqrt{-\hat g}$, we get eventually (after some algebra) the following linear equation in partial derivatives
$$
\varphi\,  \frac{\partial \varepsilon }{\partial \varphi }+3n\, \frac{\partial \varepsilon }{\partial n}=4\, \varepsilon. 
$$
The general solution of this equation has the form 
$$\varepsilon =F\left ( x \right )\, \varphi ^{4}, \quad x = \frac{n}{\varphi ^{3}}$$
with $F(0)=\sigma$. Surely, the structure of the $\varepsilon_1$-term is the same
$$
\varepsilon_1 =F_1\left ( x \right )\, \varphi ^{4},
$$
with $F_1(0) = 0$, since the $\varphi^4$-term is already present in the conservation law.
\par The Lagrange multipliers themselves are assumed conformal invariant.
\section{Equations of motion} 
\par It is convenient to divide the action integral into two parts, $S_m[n]$ and $S_m[\varphi]$:
$$S_m=S_m[n]+S_m[\varphi]$$
\begin{eqnarray}
    S_m[n] &=& -\int F(x)\varphi^4\sqrt{-g}d^4x+\int\lambda_0(u^\sigma u_\sigma-1)\sqrt{-g}d^4x+\\
    &&+\int\lambda_1((nu^\sigma)_{;\sigma}-F_1(x)\varphi^4)\sqrt{-g}d^4x+\int\lambda_2X_{,\sigma}u^\sigma\sqrt{-g}d^4x, \nonumber
\end{eqnarray}
$$S_m[\varphi]=-\int \lambda_1(\alpha C^2+\beta(\varphi\varphi^\sigma_{;\sigma}-\frac{1}{6}\varphi^2R+\Lambda\varphi^4))\sqrt{-g}d^4x,$$
$$x=\frac{n}{\varphi^3}.$$
We ignore here the dependence of function F on the auxiliary dynamical variable $X$, because its variation gives us the equation for the Lagrange multiplier $\lambda_2$ which of no use in cosmology. For the same reason we will put $C^2=0$ in the action integral because $C\delta C=0$.
\par The variation of the particle number density, $\delta n$, gives us the following equation:
$$
-\frac{dF}{dx}\varphi-\lambda_{1,\sigma}u^\sigma-\lambda_1\frac{dF_1}{dx}\varphi = 0,
$$
surely, we integrated by parts the term $\lambda_1(nu^\sigma)_{;\sigma}\sqrt{-g} = \lambda_1(nu^\sigma\sqrt{-g})_{,\sigma}$ and put zero the corresponding surface term. In the same way we obtained the result of the variation of the vector field, $\delta u^\mu$,
$$
2\lambda_0u_\mu-\lambda_{1,\mu}n+\lambda_2X_{,\mu} = 0.
$$
By the variation of $\delta X$ one gets
$$\frac{\partial \varepsilon}{\partial X}-(\lambda_2u^\sigma)_{;\sigma} = 0.$$
The constraints are
\begin{equation*}
 \begin{cases}
   u^\sigma u_\sigma - 1 = 0 \\
   (nu^\sigma)_{;\sigma}-\Phi=0 \\
   X_{,\sigma}u^\sigma = 0
 \end{cases}
\end{equation*}
By contracting the second (vector) equation, using constraints and the first (scalar) equation, we calculate the Lagrange multiplier $\lambda_0$:
$$
2\lambda_0 = -x\left( \frac{dF}{dx}+\lambda_1\frac{dF_1}{dx} \right)\varphi^4.
$$
The energy-momentum tensor $T^{\mu\nu}$ is defined by the following relation:
$$\delta = -\frac{1}{2}\int T^{\mu\nu}(\delta g_{\mu\nu})\sqrt{-g}d^4x.$$
Its ''hydrodynamical'' part, $ T^{\mu\nu}[n]$, equals 
$$
T^{\mu\nu}[n] = \varepsilon g^{\mu\nu}-2\lambda_0 u^\mu u^\nu + g^{\mu\nu}(n\, \lambda_{1,\sigma}u^\sigma+\lambda_1F_1(x)\varphi^4).
$$
Substituting for $\lambda_0$ the expression found above gives us
\begin{eqnarray}
    T^{\mu\nu}[n] &=& x\varphi^4\left(\frac{dF}{dx}+\lambda_1\frac{dF_1}{dx}\right)u^\mu u^\nu-\\
    &&- \varphi^4\left(x\frac{dF}{dx}-F+\lambda_1\left(x\frac{dF_1}{dx}-F_1\right)\right)g^{\mu\nu}.\nonumber
\end{eqnarray}
If $F_1(x) = 0$, then, remembering that $\varepsilon = F(x)\varphi^4$ and introducing the hydrodynamical pressure $p = n\frac{\partial\varepsilon}{\partial n}-\varepsilon$, one recovers the familiar hydrodynamical energy-momentum tensor
$$
T^{\mu\nu}_{hydro} = (\varepsilon + p)u^\mu u^\nu
 - pg^{\mu\nu}.
$$
\par The variation of the scalar field, $\delta\varphi$, gives the following equation (surely, after integrating by parts the term $\lambda_1\varphi\varphi^\sigma_{;\sigma}\sqrt{-g}$ and putting zero the corresponding surface term):
\begin{eqnarray}
    \left(3x\frac{dF}{dx}-4F\right)\varphi^3+\lambda_1\left(3x\frac{dF_1}{dx}-4F_1\right)\varphi^3 =\\
    = \beta\left(\lambda_1\varphi^\sigma_{;\sigma} + (\lambda_1\varphi)^{;\sigma}_{;\sigma}+4\lambda_1\Lambda\varphi^3-\frac{1}{3}\lambda_1\varphi R\right). \nonumber
\end{eqnarray}
At last, the remaining part of the energy-momentum tensor, $T^{\mu\nu}[\varphi]$ equals

\begin{eqnarray}
    T^{\mu\nu}[\varphi] = \beta\left( (\lambda_1\varphi)^{;\mu}\varphi^\nu + (\lambda_1\varphi)^{;\nu}\varphi^\mu - ((\lambda_1\varphi)_{,\sigma}\varphi^\sigma - \lambda_1\Lambda\varphi^4)g^{\mu\nu} \right)\\
    + \frac{\beta}{3}\left( \lambda_1\varphi^2\left(R^{\mu\nu} - \frac{1}{2}Rg^{\mu\nu}\right) -(\lambda_1\varphi^2)^{;\mu;\nu} + (\lambda_1\varphi^2)^{;\sigma}_{;\sigma}g^{\mu\nu}\right).\nonumber
\end{eqnarray}
Because of the induced gravity,
$$
T^{\mu\nu}[n] + T^{\mu\nu}[\varphi] = 0.
$$
It is not difficult to show that the zero value of the trace of the energy-momentum tensor, $T = T[n] + T[\varphi]$, is the consequence of the equation for the scalar field $\varphi$.
\par In the comoving Robinson-Walker coordinate system with the cosmological time $t$, the complete set of the equation looks as follows:
$$
    \dot{\lambda}_1+\lambda_1\frac{dF_1}{dx}\varphi +\frac{dF}{dx}\varphi = 0,
$$
$$
    \frac{1}{a^3}\frac{d}{dt}(na^3)=\Phi = \beta\left( \frac{\varphi}{a^3}\frac{d}{dt}(\varphi a^3) + \varphi^2\left( \frac{\Ddot{a}}{a}+\frac{\dot{a}^2+k}{a^2}+\Lambda\varphi^4\right) + F_1(x)\varphi^4\right),
$$
\begin{eqnarray}\nonumber
\varphi \ddot{\lambda }_{1}+\dot{\lambda }_{1}\left ( 3\varphi \, \frac{\dot{a}}{a}+2\dot{\varphi } \right )+\lambda _{1}\left ( 2\ddot{\varphi }+6\, \frac{\dot{a}}{a}\dot{\varphi }+4\Lambda \, \varphi ^{3 }+2 \varphi\, \left(\frac{\Ddot{a}}{a}+\frac{\dot{a}^2+k}{a^2}\right)   \right )+\\+\lambda_{1}\,\varphi^3\left(4  F_1-3x\, \frac{dF_1}{dx} \right)=-\frac{\varphi^3}{\beta }\,\left(4 F-3x \frac{dF}{dx} \right),   \nonumber
\end{eqnarray}

\begin{eqnarray} \nonumber
\left(F(x)+\lambda_1F_1(x)\right)\varphi^4+\beta\dot{\lambda}_1\varphi\left(\dot{\varphi}+\frac{\dot{a}}{a}\varphi\right)+\beta\lambda_1\left(\dot{\varphi}^2+2\frac{\dot{a}}{a}\dot{\varphi}\varphi+\Lambda\varphi^4+\varphi^2\frac{\dot{a}^2+k}{a^2}\right) = 0 .
\end{eqnarray} 
Here we used explicitly the relations for the curvature
$$
R^0_0=-3\frac{\Ddot{a}}{a},
$$
$$
R=-6\left(\frac{\Ddot{a}}{a}+\frac{\dot{a}^2+k}{a^2}\right).
$$
Not all of these equations are independent - there is the hidden conservativity of the Einstein tensor $G_{\mu\nu}=R_{\mu\nu}-\frac{1}{2}R g_{\mu\nu}$.
\par It is very useful and instructive to rewrite the above set of equations in terms of the conformal invariant functions and conformal time (which is also conformal invariant). Two conformal functions are already at hand, they are $x=\frac{n}{\varphi^3}$ and $\lambda_1(\eta)$. For the third one we will choose $f=\varphi a$.
The result is:
\begin{eqnarray}
 \nonumber   \frac{1}{f}\frac{d\lambda_1}{d\eta}+\lambda_1 \, \frac{dF_1}{dx}+\frac{dF}{dx}=0,\\
\nonumber    \frac{d(xf^3)}{d\eta}=\beta\left(f\frac{d^2f}{d\eta^2}+kf^2+\Lambda f^4\right)+F_1(x)f^4,\\
\nonumber   f^3\left(\left(3x\frac{dF}{dx}-4F\right)+\lambda_1\left(3x\frac{dF_1}{dx}-4F_1\right)\right)=\\=\beta\left(2\lambda_1 \left(\frac{d^2f}{d\eta^2}+k f+2\Lambda f^3 \right)+f\frac{d^2\lambda_1}{d\eta^2}+2\frac{d \lambda_1}{d \eta}\frac{df}{d\eta}\right),\\
\nonumber   f^4(F(x)+\lambda_1F_1(x))+\beta\left(\frac{d(\lambda_1f)}{d\eta}\frac{df}{d\eta}+\lambda_1kf^2+\lambda_1\Lambda f^4\right)=0.
\end{eqnarray}
The third equation can be reproduced by differentiating of the fourth equation and making use of the first and second ones. Surely, this is the consequence of the conservativity of the Einstein tensor. Thus, we have 3 equations for 3 conformal invariant functions, $\lambda_1(\eta)$, $x(\eta)$ and $f(\eta)$.
\par The scale factor $a(\eta)$ does not enter these equation as it should be, because it plays, actually, the role of the conformal factor $\Omega$. Therefore, the only trace of the geometry is the type of the cosmological model encoded in the value of $k$, namely, $k=1$ for the closed universe, while $k=0$ for the spatially flat and $k=-1$ for the open ones. In order to obtain the specific $a(\eta)$, we should impose the gauge fixing condition.

\section{Gravitating mirages}
\par In this section we would like to describe the quite new phenomenon - "gravitating mirages" \cite{Ber2}. They originated from the $F_1(x)$-term in the creation law and enter the energy-momentum tensor, where they multiplied by the Lagrange multiplier $\lambda_1$. Thus, they gravitate. The function $F_1(x)$ depends on the number density of the already created particles $(x=\frac{n}{\varphi^3})$ but it does not describe them - this is done by the function $F(x)$ whose structure can be quite different from that of $F_1(x)$ (for instance, the real particle consists solely of dust, $F(x) \propto x$, while $F_1(x) \propto x^\frac{4}{3}$, i.e., represents the thermal bath). Thus, $F_1(x)$ describes something that is not the matter at all, such entities can be called "the gravitating mirages".
\par Actually, $F_1(x)$ describes the back reaction of the very process of particle creation on the spacetime geometry. It constitutes the invisible part of the whole gravitating system. It is worthwhile to note two important examples. First of them is described by the linear part of $F_1(x)$: if it is nonzero, then there should exist the invisible dust in the universe what can be interpreted as the dark matter (it is dark because it is not the matter). The second example is the invisible thermal bath described by the term $~x^\frac{4}{3}$ in $F_1(x)$: its existence may cause the discrepancy between the visible homogeneous and isotropic microwave background and the total amount of the gravitating total thermal bath.

\section{Gauge fixing}
\par Up to now we did not specify the gauge. The role of the conformal factor is played by the conformal factor is played by the scale factor $a(\eta)$ (as a function of the conformal time). Evidently, one can put $a=1$ everywhere except at $a=0$, which is not, actually, the point but the null surface. In such a gauge all the above equations remain, surely, the same with $f=\varphi a \xrightarrow{} \varphi$.
\par Another interesting choice is to put
$$
\dot{\lambda}_1=\frac{d\lambda_1}{dt}=\frac{1}{a}\frac{d\lambda_1}{d\eta}=\pm1,
$$
$$
\lambda_1=\pm(t-t_0),
$$
where t is the cosmological time in this very gauge. For the function $\lambda_1(\eta)$ one has
$$
\lambda_1(\eta)=\pm\int\limits_{\eta_0}^\eta a(\eta_1)d\eta_1,
$$
what reveals the nonlocal nature of the Lagrange multiplier $\lambda_1$ and indicates its relationship with the nonlocal part of the trace anomaly (which, in turn, is responsible for the particle production). Note that the use of the cosmological time $t$ makes the set of the cosmological equations time-dependent and enters the scale factor $a(t)$ into play. Thus, we have 3 equations for 3 dynamical variables, $x=\frac{n}{f^3}$, $f=\varphi a$ and $a$ as the functions of t.
\par The most intriguing gauge fixing is 
$$
\beta\lambda_1\varphi^2=-\frac{3}{\kappa}, \quad \kappa=\frac{8\pi G}{c^4},
$$
where G is the Newtonian constant, ant c is the speed of light. First of all, with such a choice we restore the cosmological Friedmann equations (because of $C^2=0$), what allows us to compare the results with the observation, interpretation of wich was based on General Relativity. Second, we are able to eliminate the Lagrange multiplier $\lambda_1$ entirely. The appropriate name for this, obviously, "the GR-gauge". The $\begin{pmatrix}
0\\
0
\end{pmatrix}$-Friedmann equation reads now as follows
$$
\frac{\dot{a}^2+k}{a^2} = \frac{\kappa}{3} \, F(x)\varphi^4-\frac{1}{\beta} \, F_1(x)\varphi^2+\frac{\dot{\varphi}^2}{\varphi^2}-\Lambda\varphi^2,
$$
$\left(x=\frac{n}{\varphi^3}\right)$. By introducing
$$\chi=\chi_0 \, ln\left|\frac{\varphi}{\varphi_0}\right|,
$$
one gets the energy-momentum tensor for the real (not virtual) scalar field with the correct sign of the kinetic term. This reminds us the $f(R)$-gravity in the so called Einstein frame. The direct calculation show that 
$$
f(R)\propto R^\frac{3}{2}.
$$

\section{Vacua}
\par By "vacua" we understand the solutions with no particles and no particle creation. Thus,
$$
x=0, \quad \frac{dx}{d \eta}=0 .
$$
Let us introduce the constants: $F(0)=\sigma,\quad \frac{dF}{dx}(0)=\mu_1,\quad  \frac{dF_1}{dx}(0)=\gamma_1$, and we remember that $F_1(0)=0$. Our set of equations becomes:
$$
\frac{d \lambda_1}{d \eta}+f \left(\mu_1+\lambda_1 \, \gamma_1 \right)=0,
$$
$$
 f\left\{ \frac{\mathrm{d}^{2}f }{\mathrm{d} \eta ^{2}}+k \, f+\Lambda \, f^3\right\}=0,
$$
$$
\sigma f^{4}+\beta \left\{ \frac{\mathrm{d} \left ( \lambda_1 f \right )}{\mathrm{d} \eta }\frac{\mathrm{d} f}{\mathrm{d} \eta }+\lambda_1 \, k f^2+\lambda_1\, \Lambda f^4\right\}=0,
$$
where $k=0, \pm 1$ denotes the type of the cosmological model. We did not include in this set the equation obtained by variation $\delta \varphi$, since it is the differential consequence of all other equations due to the conservativity of the Einstein tensor. Note, that we have 3 equations for only 2 functions, $\lambda_1(\eta)$ and $f(\eta)$, so set of the equations is overdetermined. The second equation is splittеd naturally into two branches: either $f \equiv 0$, or
$$
\frac{\mathrm{d}^{2}f }{\mathrm{d} \eta ^{2}}+k \, f+\Lambda \, f^3=0,
$$
the first integral of which can be written in the form
$$
\left ( \frac{\mathrm{d} f}{\mathrm{d} \eta } \right )^{2}+U(f)=0, \quad U(f)=kf^2+\frac{1}{2}\Lambda f^4-C_0, \quad C_0=const.
$$
\par There are only three different situations:
\par $\#1$.
$$
f(\eta)\equiv 0, \quad \lambda_1=const.
$$
No way to determine the geometry (i.e., $a(\eta)$ ).
\par  $\#2$.
$$
f=f_0=const, \quad f^{2}_{0}=-\frac{k}{\Lambda}, \quad \sigma=0,
$$
$$
\lambda_1(\eta )=\left ( \lambda_1(0)+\frac{\mu_1}{\gamma_1} \right )e^{-f_0\gamma_1 \eta }-\frac{\mu_1}{\gamma_1}.
$$
Note, that for $k=0$ it coincides with the vacuum  $\#1$. In the GR-gauge $\left ( \beta \lambda_1 \varphi^2=-\frac{3}{\kappa },\; \kappa =\frac{8 \pi G}{c^4}, \; f=\varphi a \right )$ one has
$$
a^{2}=-\frac{\kappa \beta }{3}f_{0}^{2}\, \lambda_1(\eta ), \quad (\beta \lambda_1<0). 
$$
\par $\#3$.
$$
\lambda_1=-\frac{\mu_1}{\gamma_1}, \quad \sigma=\frac{1}{2}\frac{\beta \mu_1}{\gamma_1}\Lambda, \quad C_0=0.
$$
Note, that this does not require the special value for $\lambda_1$. On the contrary, it is the value of $\lambda_1$ that determines the characteristics of the creating particles. The special value of $\sigma$ means the choice of the solution (since $\sigma$ is just the constant of integration). For the function $f(\eta)$ we have now the following equation $(f\not\equiv 0)$:
$$
\left ( \frac{\mathrm{d} f}{\mathrm{d} \eta } \right )^{2}+kf^2+\frac{1}{2}\Lambda f^4=0.
$$
It can be solved explicitly for different values of $k$:
\begin{itemize}
    \item $k=0,\; \Lambda<0: \quad f^2=-\frac{2}{\Lambda}\frac{1}{(\eta-\eta_0)^2}$,
    \item $k=-1,\; \Lambda>0: f^2=\frac{2}{\Lambda}\frac{1}{ch^2(\eta-\eta_0)}$,
     \item $k=-1,\; \Lambda<0: f^2=-\frac{2}{\Lambda}\frac{1}{sh^2(\eta-\eta_0)}$,
     \item $k=1,\; \Lambda<0:  f^2=-\frac{2}{\Lambda}\frac{1}{sin^2(\eta-\eta_0)}$.
\end{itemize}
The scale factor in GR-gauge equals:
$$
a^2=\frac{\mu _{1}}{\gamma _{1}}\frac{\kappa \beta }{3}f^{2}.
$$
\par Important note: there cannot be classical transitions between different vacua.
\par Let us consider different cases.
\subsection{\texorpdfstring{$\Lambda>0,\; k=0, \; C_0 >0$}{TEXT}}
\begin{figure}[hbt!]
	\includegraphics[width=0.5\textwidth]{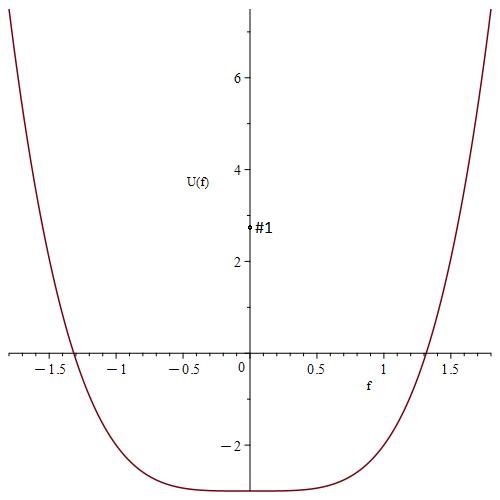}
	\caption{\small{Graph of $U(f)$ \quad $(k=0,\, \Lambda=2, \, C_0=3)$} }
	\label{Fig1}
\end{figure}
\par The vacuum $\#2$ coincides with the vacuum $\#1$. The vacuum $\#3$ is absent. Any deviation from the value $f=0$ causes the creation of particles.
\subsection{\texorpdfstring{$\Lambda>0,\; k=1, \; C_0 > 0$}{TEXT}}
\par See Fig.\ref{Fig1}.
\subsection{\texorpdfstring{$\Lambda>0,\; k=-1, \; C_0 > 0, \; f^{2}_{0}=\frac{1}{\Lambda}$}{TEXT}}

\begin{figure}[hbt!]
	\includegraphics[width=0.5\textwidth]{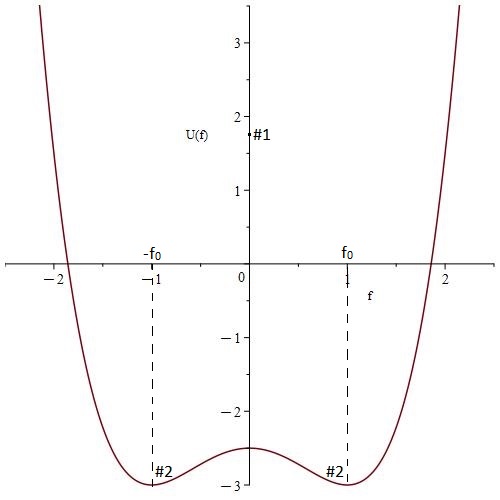}
	\caption{\small{Graph of $U(f)$ \quad $(k=-1,\, \Lambda=1, \, C_0=2.5)$} }
	\label{Fig2}
\end{figure}
\par Fig.\ref{Fig2} shows that two vacua are possible to exist, $\#1$ and $\#2$. If $C_0<0$, the maximum of the curve should be placed on the upper half of the plane.
\subsection{\texorpdfstring{$\Lambda<0,\; k=0, \; C_0 > 0$}{TEXT}}
\begin{figure}[hbt!]
	\includegraphics[width=0.5\textwidth]{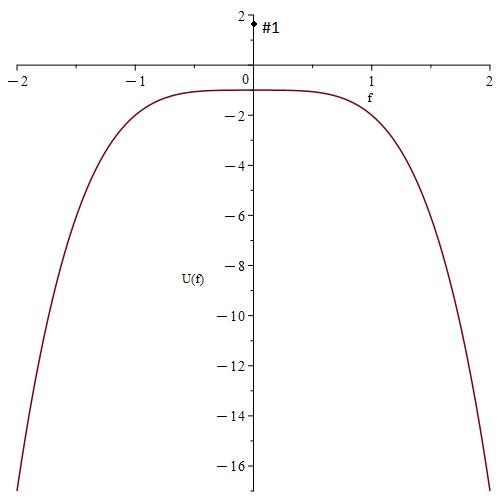}
	\caption{\small{Graph of $U(f)$ \quad $(k=0,\, \Lambda=-2, \, C_0=1)$} }
	\label{Fig3}
\end{figure}
\par For the Fig. \ref{Fig3} the maximum should lie at $U>0$ if $C_0<0$.
\subsection{\texorpdfstring{$\Lambda<0,\; k=1, \; C_0 > 0$}{TEXT}}
\begin{figure}[hbt!]
	\includegraphics[width=0.5\textwidth]{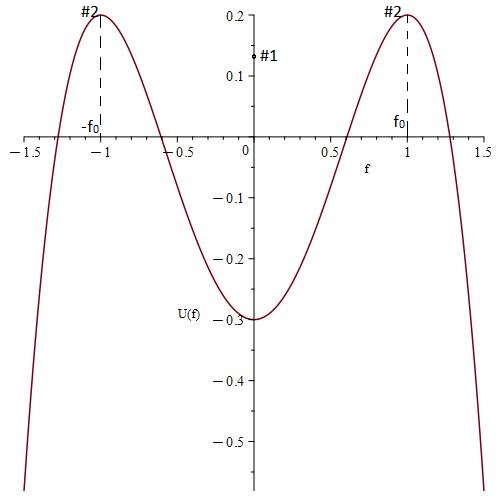}
	\caption{\small{Graph of $U(f)$ \quad $(k=1,\, \Lambda=-1, \, C_0=0.3)$} }
	\label{Fig4}
\end{figure}
\par For the case depicted in the Fig. \ref{Fig4} two different vacua are possible, $\#1$ and $\#2$.

\par At last, $C_0=0$. There appears the vacuum $\#3$. We will not show here the case $C_0=0, \; \Lambda>0, \; k=0,\, 1$, where only the vacuum $\#1$ exists.
\subsection{\texorpdfstring{$\Lambda>0,\; k=-1, \; C_0=0$}{TEXT}}
\begin{figure}[hbt!]
	\includegraphics[width=0.5\textwidth]{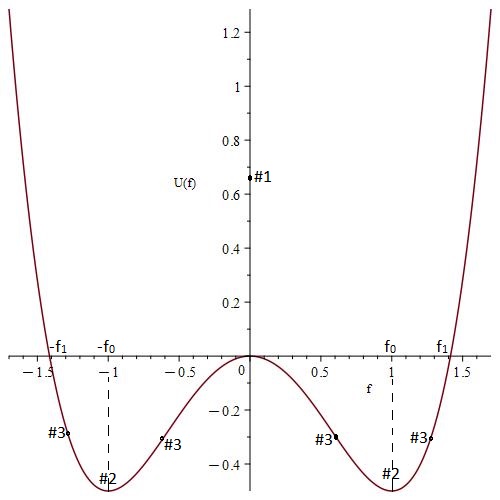}
	\caption{\small{Graph of $U(f)$ \quad $(k=-1,\, \Lambda=1, \, C_0=0)$} }
	\label{Fig5}
\end{figure}
\par In the Fig.\ref{Fig5} all three types of vacua is present. The vacuum $\#3$ may emerge at any point in the regions $-f_{1} \leqslant f <0$ or $0<f \leqslant  f_1 $ $\left(f_1=\sqrt{\frac{2}{\Lambda}}\right)$. Since in this case
$$
f^2=\frac{2}{\Lambda} \, \frac{1}{ch^2(\eta-\eta_0)},
$$
then, at $\eta=\eta_0$ one obtains $f^2=\frac{2}{\Lambda}$, and the evolution towards $f=0$ takes the infinite time. Correspondingly, the scale factor $a(\eta)$ in the GR-gauge starts from the maximal value at $f=f_1$ and tends to zero for $f \to 0$ (and vice verse).
\subsection{\texorpdfstring{$\Lambda<0,\; k=0,\, -1, \; C_0 = 0$}{TEXT}}
\par See Fig. \ref{Fig3}.
\subsection{\texorpdfstring{$\Lambda<0,\; k=1, \; C_0 = 0$}{TEXT}}
\begin{figure}[hbt!]
	\includegraphics[width=0.5\textwidth]{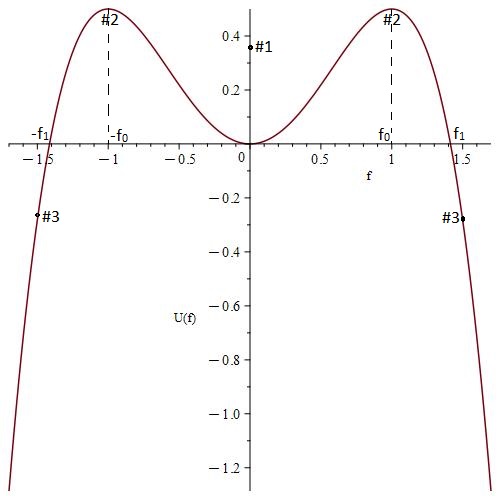}
	\caption{\small{Graph of $U(f)$ \quad $(k=1,\, \Lambda=-1, \, C_0=0)$} }
	\label{Fig6}
\end{figure}
\par The last case depicted in the Fig. \ref{Fig6}. Both vacua $\# 1$ and $\# 2$ undergoes the phase transition to the particle creation for slightest deviation of the corresponding value of $f$. The situation with the vacuum $\# 3$ is different. It may emerge at any point $|f| \leqslant |f_1|$ and evolve according to the law
$$
f^2=\frac{2}{|\Lambda|}\frac{1}{sin^2(\eta-\eta_0)}.
$$
Correspondingly, in the GR-gauge the scale factor increases from some minimal value to the infinity (and vice verse). Almost de Sitter space-time! Besides, the vacuum $\#1$ may undergo the quantum tunneling into the vacuum $\#3$ with $\lambda_1=const$, thus determining the value of $\mu_1$ (the value of $\gamma_1$ is fixed in the Lagrangian). Such a process begins at the past infinity of the imaginary time $i \eta$. Thus, since the period of the imaginary time is infinite, the temperature after this transition is zero.

\section{Discussion and Conclusion}
\par In this paper we considered the phenomenological description of the particle creation process in the framework of the induced gravity. The created particles are described by the perfect fluid, and the action integral of such a fluid in the Eulerian dynamical variables is modified by replacing the (usual) particle number conservation law by the explicit particle creation law. It appears that this creation law function is conformal invariant ( when multiplied by the square root of the metric determinant). Moreover, the induced gravity requirement leads to the conformal invariance of the variation of the constrained matter integral (of course, up to the surface term which is zero on the classical trajectories). Thus, the conformal invariance play the important role in our model.
\par The proposed phenomenological description of the particle creation allows for the accounting of the back reaction on the space-time metric not only of the already created particles but also the very process of the creation, which is, of course, of the quantum nature but influences classically. Therefore, we implicitly include into the consideration the well known trace anomaly.
\par We confined ourselves to the cosmological application assuming that the created particles are quanta of the scalar field. It appeared that the accounting for the back reaction makes it possible to create particles in the case of the homogeneous and isotropic cosmology unlike the famous results obtained for the background consideration \cite{Cr1,Alstar1,Cr3}.
\par We include into the creation law not only the square of the Weyl tensor and the famous combination of the scalar field with the curvature scalar which are allowed by the conformal invariance, but also the function of the number density of the already created particles. The latter, being the real (not virtual) quanta of the scalar field, may also cause both the creation and annihilation processes. To our surprise, these quite new terms enter the energy-momentum tensor. They are not the real matter and it is impossible to "touch" them, but they are gravitating. We called them "the gravitating mirages". They may constitute the invisible part of the Universe, such as the dark matter and the part of the cosmological microwave background. Their contributions should be taken into account, because they are gravitating. 
\par We also managed to rewrite the cosmological field equations using only conformal invariant dynamical variables. Since the scale factor in the homogeneous and isotropic cosmological models may serve as the conformal factor, it disappears from the obtained set of equations. The only way to reveal the scale factor is to impose some gauge fixing condition. The most "physical" of them is that one that restores the General Relativity. We called such a gauge fixing "the GR-gauge". In such a case one can compare the theoretical results with the observation in a direct way. It appeared that in the GR-gauge our model is equivalent to the so called $f(R)$-gravitational models for $f \propto R^{\frac{3}{2}}$.

\end{document}